\pdfoutput=1                            
\def\lcversion{}					    

\documentclass[noversion,squeezed]{cdcarticle}

\def\MODE{2}

\usepackage{cite}                   
\usepackage{math}                   
\usepackage[utf8]{inputenc}			
\usepackage[english]{babel}			
\usepackage[hidelinks]{hyperref}    

\if\MODE2\else
\usepackage{amsthm}
\newtheorem{thm}{Theorem}
\newtheorem{problem}[thm]{Problem}
\newtheorem{lem}[thm]{Lemma}
\newtheorem{prop}[thm]{Proposition}
\newtheorem{rem}[thm]{Remark}
\newtheorem{defn}[thm]{Definition}
\newtheorem{cor}[thm]{Corollary}
\newtheorem{assumption}[thm]{Assumption}
\def\qed{\rule[0pt]{5pt}{5pt}\par\medskip}
\renewcommand{\qedhere}{\hfill ~\qed}

\fi

\usepackage{bbm}                    
\usepackage{textcomp}				
\usepackage{graphicx}
\graphicspath{{./graphics/}}

\newcommand{\U}{\mathcal{U}}
\newcommand{\V}{\mathcal{V}}
\newcommand{\W}{\mathcal{W}}	
\newcommand{\G}{\mathcal{G}}
\newcommand{\Pp}{\mathcal{P}}

\newcommand{\K}{\mathcal{K}}

\newcommand{\Stau}{\mathcal{S}_{\tau}}

\usepackage{tikz}
\usepackage{amsmath}
\usetikzlibrary{calc,positioning}
\usepackage{amssymb}
\if\MODE3
\fi

\usepackage{ntheorem}
\usepackage[capitalise,nameinlink]{cleveref}
\usetikzlibrary{shapes.geometric}

\newtheorem{thm}{Theorem}

\newtheorem{lem}[thm]{Lemma}

\crefname{thm}{Theorem}{Theorems}
\crefname{problem}{Problem}{Problems}
\crefname{lem}{Lemma}{Lemmas}
\crefname{prop}{Proposition}{Propositions}
\crefname{rem}{Remark}{Remarks}
\crefname{defn}{Definition}{Definitions}
\crefname{cor}{Corollary}{Corollaries}
\crefname{assumption}{Assumption}{Assumptions}

\begin{document}
\title{Optimal decentralized wavelength control in light sources for lithography}

\if\MODE1
\author{Mruganka Kashyap}
\note{Submitted to ACC 2025}
\fi

\if\MODE2
\author{Mruganka Kashyap}
\note{Submitted to IEEE American Control Conference 2025}
\fi

\if\MODE3
\def\BibTeX{{\rm B\kern-.05em{\sc i\kern-.025em b}\kern-.08em
    T\kern-.1667em\lower.7ex\hbox{E}\kern-.125emX}}
\markboth{\journalname, VOL. XX, NO. XX, XXXX 2023}
{Kashyap \MakeLowercase{\textit{et al.}}: Preparation of Papers for IEEE Control Systems Letters (2023)}

\author{
	Mruganka Kashyap$^{1}$ \IEEEmembership{Member, IEEE}
	\thanks{M.~Kashyap is a Senior Controls Engineer at ASML. (e-mail: mruganka.kashyap@asml.com)}
}
\fi

\maketitle
\thispagestyle{empty}

\if\MODE3
\footnotetext[1]{M.~Kashyap is a Senior Controls Engineer with the Light Control Group at Cymer, LLC, an ASML company, San Diego, CA 92069, USA. (e-mail: mruganka.kashyap@asml.com).
	
 }
\fi


\begin{abstract}
	Pulsed light sources are a critical component of modern lithography, with fine light beam wavelength control paramount for wafer etching accuracy. We study optimal wavelength control by casting it as a decentralized linear quadratic Gaussian (LQG) problem in presence of time-delays. In particular, we consider the multi-optics module (optics and actuators) used for generating the requisite wavelength in light sources as cooperatively interacting systems defined over a directed acyclic graph (DAG). We show that any measurement and other continuous time-delays can be exactly compensated, and the resulting optimal controller implementation at the individual optics-level outperforms any existing wavelength control techniques.	 
\end{abstract}


\section{Introduction}\label{sec:intro}

The
current information economy including the advent of Artificial Intelligence (AI) based generative modeling, and internet of things (IoT) has its foundations embedded in over half a century of advancements in the semiconductor industry, which has pushed the boundaries of Moore's Law~\cite{Moore}. The biggest enabler of this is the photolithography process, which is currently the industry standard for manufacturing microchips. The photolithography process constitutes two main components: the wafer scanner, and a light source~\cite{Levinson}.

Wafer scanners consist of an intercombination of highly complex mechatronic systems, which combine high wafer throughput with high precision for wafer-etching. The scanning process is performed using a sequence of concatenated point-to-point motions, with the tracking specifications of the scanner’s motion systems lying in the (sub-)nanometer ranges~\cite{Levinson,kevin_tutorial}.  On the other hand, the light source, also known in industry as the laser, is a nonlinear, Multi-Input Multi-Output (MIMO) system. Light is
generated in the form of bursts of pulses at several kHz, considered the repetition rate of the laser. These pulses illuminate a mask and expose the photo-resistive material on silicon wafers~\cite{Levinson}. Each burst normally corresponds to one die on the wafer, and is followed by a quiescent interval, referred to as the inter-burst interval, which corresponds to moving an adjacent die into position for exposure. Since no bursts are fired in this interval, no measurements of the light are available in this period. 

\begin{figure}
    \centering
    \includegraphics[width=0.6\linewidth]{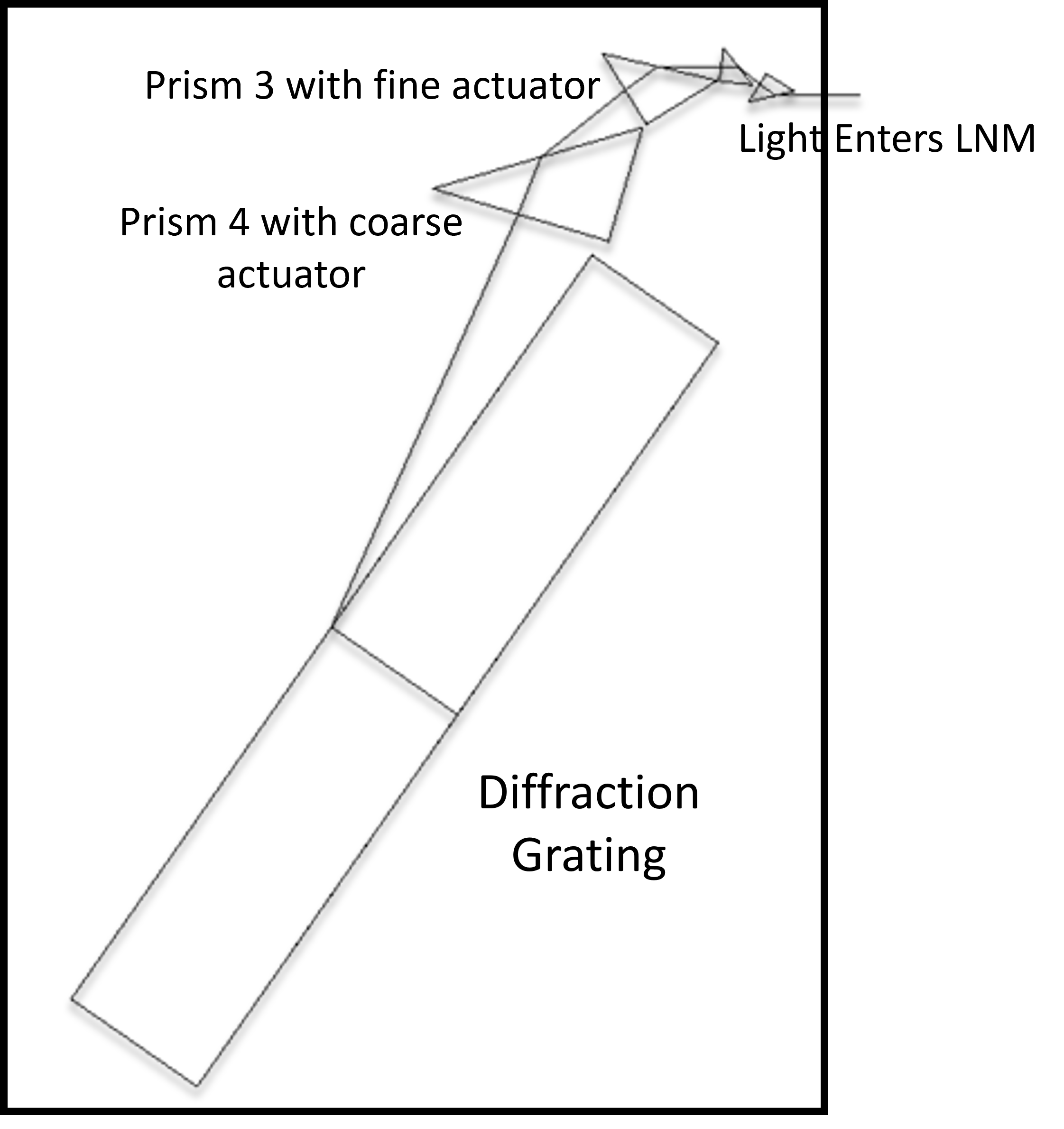}
    \caption{Schematic of a Line Narrowing Module (LNM) based on multi-prism dispersion theory. An LNM is an example of a combination of optics used for wavelength and bandwidth control in light sources.}
    \label{fig:LNM}
\end{figure}

The three important performance specifications of the light include stability in energy, wavelength, and bandwidth, which directly impact the on chip resolution metric known as the Critical Dimension ($\textup{CD}\defeq k \frac{\lambda}{\textup{NA}}$), where $k$ represents the $k$-factor for a given process, $\lambda$ the wavelength, and $\textup{NA} > 0$ the numerical aperture~\cite{Levinson}. The typical values for the so-called DUV (deep ultraviolet) technology in lithography are $k = 0.25$, $\lambda = 193\; \textup{nm}$, and $\textup{NA}=1.35$ leading to $38\; \textup{nm}$ line widths for $\textup{ArF}$ gas light sources~\cite{french2005imaging,rothschild1988review}. As such, the wavelength of the laser directly determines the size of the feature being printed. Wavelength control is achieved in modern light sources using a combination of optics that have reflective and diffractive properties, including multiple prisms, deformable mirrors, and lenses~\cite{chang1991multiple,coldren2004tunable,kevin_tutorial,mroziewicz2008external,Paschottaultraviolet_optics}. For instance, state-of-the-art DUV light sources at Cymer, LLC (an ASML company), an industry leader in light sources, have a Line Narrowing Module (LNM) that controls the light source wavelength in its Master Oscillator (MO) chamber, where the seed light is generated~\cite{blumenstock2005evolution}. An LNM leverages the concepts of multiple-prism arrays and multiple-prism dispersion to control bandwidth and wavelength of a light source~\cite{newton1952opticks,duarte1998long}.~\cref{fig:LNM} represents a schematic of multiple prisms cooperatively performing wavelength control for a light source.

In ~\cref{fig:LNM}, light enters the LNM, and it travels through four different prisms before the diffraction grating disperses the incoming light. The directions of reflection are dependent upon the wavelength of the photons in the light beam. The wavelength in the output light is a function of the angle of incidence on the diffraction grating, prism geometry, prism refractive index, and number of prisms~\cite{duarte1998long,duarte1983generalized}.
While all four prisms have an effect on the wavelength, the prism closest to the grating (Prism $4$) has a larger gain from the prism angle to wavelength, followed by the previous one (Prism $3$)~\cite{kevin_tutorial}. The wavelength is modified by the actuation of
mechanical devices like solenoids, piezoelectric transducers (PZT), stepper motors, which are affixed to the various optics and actuate their corresponding positions/angles. For instance, Prism $4$ in~\cref{fig:LNM} is driven by a stepper motor and serves as a coarse wavelength control, while the Prism $3$ is actuated by a PZT, acting as a fine wavelength control. 

In this paper, we consider a two-optics module system for wavelength control for the sake of simplicity. Without loss of generality, we will assume one of the instrument is actuated by a PZT, while the other by a stepper motor. This assumption allows us to cover most of the standard actuators used for wavelength control in industry. We will use the term `prism' to refer to any individual optics used for wavelength control in the remainder of the paper.

In \cref{sec:Decen} we review more recent work on decentralized linear quadratic Gaussian (LQG) synthesis, which form the basis for optimal cooperative wavelength control synthesis. In \cref{sec:problem} we provide the problem setup for optimal decentralized wavelength control and we present our main results in \cref{sec:main}. In \cref{sec:discussion,sec:conclufuture} we present alternate implementations, interpretations, and future work.

\section{Decentralized LQG control}\label{sec:Decen}

Dynamically decoupled decentralized systems are examples of cooperative control systems where each sub-system has independent dynamics but the sub-systems' controllers share information via a
communication topology to optimize a common, global cost. The communication network is abstracted as a directed acyclic graph (DAG). Each system is a linear time-invariant (LTI) dynamical system of the form:
\begin{align}\label{eq:state_space}
    \dot x_i(t) &= A_{ii}x_i(t) + B_{1_{ii}} w_i(t) + B_{2_{ii}} u_i(t),\\
    y_i(t) &= C_{2_{ii}} x_i(t) + D_{21_{ii}} w_i(t),
    \label{eq:state_space_2_}
\end{align}
where $x_i(t)\in\R^{n_i}$ and $u_i(t) \in \R^{m_i}$, and $y_i(t) \in \R^{p_i}$ are the state, input, and measurement for each sub-system, and $w_i(t) \in R^{q_i}$ is exogenous standard Gaussian noise, independent across sub-systems and time. The infinite-horizon LQG problem is to find the causal policy that minimizes the quadratic cost $J$ is defined as 
\begin{equation}\label{eq:cost}
\lim_{T\to \infty}\frac{1}{T}\E_{w} \int_{0}^{T-1}\left( x(t)^\tp Q x(t) + u(t)^\tp R u(t)\right)\,\mathrm{d}t.
\end{equation}
The global matrices $A$, $B_1$, $B_2$, $C_2$, and $D_{21}$, obtained by stacking~\cref{eq:state_space,eq:state_space_2_} for each of the sub-systems, are block-diagonal in structure, and conform to the adjacency matrix of the transitive closure of the DAG. No assumptions are made on the cost matrices $Q$ and $R$, so all states and inputs may be coupled. Consider the 2-node DAG in \cref{fig:dag_example}.

\begin{figure}[ht]
	\centering
	\begin{minipage}{0.3\linewidth}
		\centering
		\begin{tikzpicture}[semithick]
			\tikzstyle{vertex} = [draw,circle,inner sep=1mm,font=\small]
			\tikzstyle{arr} = [>=latex,->]
			\def\y{0.6}
			\def\x{1.0}
			\node[vertex] (1) at (0,0) {$1$};
			\node[vertex] (2) at (\x,0) {$2$};
			\draw[arr] (1) -- (2);
		\end{tikzpicture}
	\end{minipage}%
	\begin{minipage}{0.3\linewidth}
		\[
		\mathcal{S} = \bmat{* & 0 \\
			* & * }
		\]
	\end{minipage}
    \begin{minipage}{0.4\linewidth}
    \begin{align*}
        A &\defeq \bmat{A_{11} & 0 \\ 0& A_{22}} \in \mathcal{S},\quad B_2 \defeq \bmat{B_{2_{11}} & 0 \\ 0& B_{2_{22}}} \in \mathcal{S}
    \end{align*}
		
  \end{minipage}
	\caption{Example of a 2-node directed acyclic graph (DAG). The communication graph has the structure of $\mathcal{S}$, shown on the right. The global matrices $A$ and $B_2$ belong to the set $\mathcal{S}$.}
	\label{fig:dag_example}
\end{figure}
\noindent What makes this problem decentralized is that each control signal $u_i$ only has access to the past history of the nodes, for which there exists a directed path from them to node $i$. So $u_i$ takes the form $u_1 = \mathcal{K}_1(x_1), \;\;
u_2 = \mathcal{K}_2(x_1,x_2)$, where the $\mathcal{K}_i$ are causal maps. 

In general, decentralized LQG problems are intractable, and need not have linear optimal control policies \cite{witsenhausen1968counterexample,wonham1968separation}. However, LQG problems
with partially nested information (plants and controllers structured according to a DAG) have a linear optimal control policy~\cite{ho1972team} and solving for this optimal linear controller can be cast as a convex optimization problem~\cite{rotkowitz2002decentralized}. Explicit closed-form solutions have been obtained for decentralized linear quadratic regulator (LQR) problem using a state-space approach~\cite{kim2015explicit,swigart_spectral} and poset-based approach~\cite{shah2013cal}, for LQG (output-feedback) problem \cite{lessard2015optimal,kim2015explicit,tanaka2014optimal,kashyap2019explicit} and also with time delays \cite{kashyap2020agent,lamperski2015optimal,10373880}. These results hold for both continuous-time and discrete-time systems, and we will leverage the corresponding results to solve a continuous-time to discrete-time controller version of our problem. We present our problem setup in the next section.

\section{Problem Setup} \label{sec:problem}
\begin{figure}[ht]
	\centering
	\begin{tikzpicture}[>=latex, node distance=8mm,semithick]
		\node[draw, inner sep=2mm] (D) {$\textup{Prism}$};
		\node[draw, inner sep=2mm, left= of D] (L) {$\textup{Actuator}$};
        \node[draw, inner sep=2mm, below= of D] (C) {$\textup{Controller}$};
        \node[circle, draw,	fill= blue!30, inner sep=1 mm, below= of L, left = 25mm of C, rotate=0] (Z) {$\textup{D/A}$ };
        \node[circle, draw, fill= blue!30, inner sep =1mm, right = of D] (S) {$\textup{LAM}$};
  
		\draw[->] (L) -- (D);
        \draw[->] (D) --(S);
        \draw[->] (C) --(Z);
		\path (S.east) -- ++(-0.5,0) coordinate (t);
		\draw[->] (S.south) |- (C.east);
        \draw[->] (Z.north) |- (L.west);

        \node[yshift=-4mm, xshift = 12mm] at (C.east) {\footnotesize $y[t]=K^{-1}_{Oi}\lambda[t]$};
        \node[yshift=-4mm, xshift = -12mm] at (C.west) {\footnotesize $u[t]=\Delta V$};
        \node[yshift=4mm, xshift = 3mm] at (Z.north) {\footnotesize $u(t)$};
        \node[yshift=-3mm, xshift = 4mm] at (D.east) {\footnotesize $y(t)$};
        \node[yshift=0mm, xshift = 3mm] at (S.east) {\footnotesize $\lambda[t]$};
	\end{tikzpicture}
	\caption{A representation of a standard prism-actuator combination in modern-day light sources used in lithography. The system is in continuous-time, while the controller is digital, requiring sample (LAM) and hold blocks. Note that we are dealing with a continuous-time system with a discrete-time controller.}
	\label{fig:prism_actuator}
\end{figure}
The block-diagram in~\cref{fig:prism_actuator} represents a standard prism-actuator combination in a laser. The Line Analysis Module (LAM) serves as a wavelength sensor and sampler (A/D) that converts continuous-time outputs of the system into discrete measurements. These measurements are fed into a controller that generates actuation signals for the actuator, which are converted into continuous-time signals via a Zero-Order Hold (ZOH), serving as a D/A. The PZT is modeled as a second-order system (two states) along with an integrator state for reference tracking, while the plant (prism) itself is a constant gain~\cite{riggs2012rejection}. Analogous linear parameterization exists for a stepper motor and its corresponding gain~\cite{mclean1988review,kim2012microstepping,blauch1993high,shah2004field}. Note that outputs $y(t)$ of the two systems are corresponding prism positions, while we have a single measurement; the overall wavelength. There exist DC, optics gains defined from the prism positions to wavelength: $K_{OP}$ and $K_{OS}$ for Prisms $3$ and $4$ respectively~\cite{kevin_tutorial}. We define the global continuous-time system as:
\begin{align}\label{eq:state_space_plant}
    \dot x &= \left(\begin{smallmatrix}
        A_{P} & 0\\
        0 & A_{S}
    \end{smallmatrix}\right)x + \left(\begin{smallmatrix}
        B_{1_{P}} & 0\\
        0 & B_{1_{S}}
    \end{smallmatrix}\right) w + \left(\begin{smallmatrix}
        B_{2_{P}} & 0\\
        0 & B_{2_{S}}
    \end{smallmatrix}\right) u,\\
    y &= \left(\begin{smallmatrix}
        C_{2_{P}} & 0\\
        0 & C_{2_{S}}
    \end{smallmatrix}\right) x + \left(\begin{smallmatrix}
        D_{21_{P}} & 0\\
        0 & D_{21_{S}}
    \end{smallmatrix}\right) w,
    \label{eq:state_space_2}
\end{align}
where $X_S$ are state-space matrices for the stepper, while $X_P$ are those of the PZT for all $X \in \{A,B_1, B_2, C_2, D_{21}\}$. The global state-vector $x\defeq \left[x_P^\tp, x_S^\tp\right]^\tp$ is formed by stacking the states of the PZT, followed by the stepper. We form the global input $u$, output $y$, and noise $w$ in a similar manner. 

\subsection{Cost}
The goal of the multi-prism configuration is to minimize the error ($\lambda-\hat{\lambda}$) between the reference ($\lambda$) and actual ($\hat\lambda$) wavelength. In reality, this wavelength error is $\lambda-\hat{\lambda} = K_{OP} (y_{P}-y_{P_{\textup{ref}}})+ K_{OS} (y_{S}-y_{S_{\textup{ref}}})$, a combination of errors incurred from both prism positions. Without loss of generality, the cost function $J$ is defined as:
\begin{equation}\label{eq:cost_problem}
\lim_{T\to \infty}\frac{1}{T}\E_{w} \int_{0}^{T-1} \Bigl(Q\normm{x}^{2}_{2} + \sum_{i\in \{P,S\}} \rho_i \normm{u_i}^2_2\Bigr) \;\mathrm{d}t,
\end{equation}
where $Q\succeq 0$ is a weighting on the states of the overall system, $\rho_P\succ 0$ and $\rho_S\succ 0$ penalize the control signals for the PZT and the stepper respectively. The state weighting matrix $Q$ is a function of the optics gains $K_{OP}$ and $K_{OS}$, and allows for coupling between the states of the two sub-systems: PZT $+$ Prism $3$, and Stepper $+$ Prism $4$. Since the PZT is a finer actuator with a limited range, $\rho_P > \rho_S$ holds in practice.  Note that we are considering the infinite-horizon case in this paper for simplicity, with the finite horizon cost briefly discussed in \cref{subsec:finite}. 

\subsection{Time-delays and DAG}
\begin{figure}[ht]
	\centering
 \begin{minipage}{0.4\linewidth}
		\includegraphics[width=1.1\linewidth]{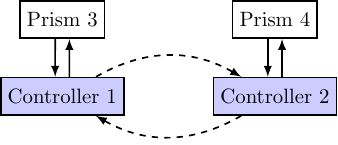}
	\end{minipage}
	\begin{minipage}{0.4\linewidth}
		\raggedleft
		\includegraphics[width=0.8\linewidth]{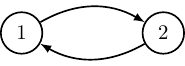}
	\end{minipage}%
	
    \caption{The 2-Prisms (with their actuators) and corresponding controllers are abstracted as 2-node directed acyclic graph (DAG). Without considering any delays, the communication graph has the structure $\mathcal{S}=\left(\begin{smallmatrix}
        * & *\\
        * & *
    \end{smallmatrix}\right)$.}
	\label{fig:prisms_dag}
\end{figure}

To the best of our knowledge, existing state-of-the-art control strategies lack communication between each other, with both prisms controlled separately in a discontinuous manner~\cite{riggs2012rejection,kevin_tutorial}. However, since both prisms simultaneously control the wavelength, we establish two-way communication between the controllers as shown in~\cref{fig:prisms_dag}.

\noindent Time-delays are incurred during computation and communication of signals, named \textit{input delays}, between the two sub-systems. Further there exists a measurement delay $\tau_1$ for both sub-systems due to the LAM processing. For this paper, we stick to constant delays and avoid complications associated with time-varying delays. Since the measurement delays are exactly the same for both sub-systems, we leverage the time-invariance property and commutativity with the plant to lump them together with the input delays. Thus, sub-system~$i$'s feedback policy (in the Laplace domain) is of the form\footnote{There is no loss of generality in assuming a linear control policy~\cite{10373880}.}
\begin{equation}\label{Kform}
u_i = e^{-s\tau_1}\K_{ii}(s)y_i + e^{-s\tau_2}\K_{ij}(s) y_j,
\end{equation}
where $i\in\{P,S\}$, $j= \{P,S\}\;\cap\; i^{\mathsf{c}}$, and $2\tau_1> \tau_2 > \tau_1$. While the rationale for $\tau_2>\tau_1$ is trivial, the $\tau_2<2\tau_1$ assumption is to satisfy the triangle inequality, i.e., information travels along the shortest path~\cite{lampHtwoDelay,rotkowitz2005characterization}. Defining $\Stau$ as the set of controllers with structure $S$ in~\cref{fig:prisms_dag} and delays $\tau_1$ in the diagonal and $\tau_2$ along cross-diagonal elements, we cast the optimal wavelength control problem as:
\begin{equation}\label{opt}
\begin{aligned}
\underset{u}{\min}
\qquad & \E_{w} \int_{0}^{T\to\infty} \Bigl(Q\normm{x}^{2}_{2} + \sum_{i\in \{P,S\}} \rho_i \normm{u_i}^2_2\Bigr)\, \mathrm{d}t \\
\subject \qquad & \K \in \Stau
\text{ and~\cref{eq:state_space_plant,eq:state_space_2} hold,}
\end{aligned}
\end{equation}
where $\K = \left(\begin{smallmatrix}
        e^{-s\tau_1}\K_{P} & e^{-s\tau_2}\K_{SP}\\
        e^{-s\tau_2}\K_{PS} & e^{-s\tau_1}\K_{S}
    \end{smallmatrix}\right)$.

In the next section, we present the solution to \cref{opt}. It is worth pointing out that there is no restriction on the time-delays and all four delay terms in $\K$ could be different from each other. The solution would still hold provided the delays satisfy the triangle inequality~\cite{rotkowitz2005characterization,spdel,lampHtwoDelay,lessard2016convexity}. 

\section{Decentralized wavelength control}\label{sec:main}
Kashyap~{\!\!\cite[Sec.~4.3]{kashyap2023thesis}} provides an explicit state-space formula for the controller and a recursive technique to handle heterogeneous delays that optimizes~\cref{opt} for $N$ sub-systems. This result will be the starting point for our work. We present a modification of this result below.

\begin{lem}{\!\!\cite[Thm.~43]{kashyap2023thesis}}\label{lem:het_decen}
    Consider the global plant $\Pp$ defined as $\left(\begin{smallmatrix}
        z\\
        y
    \end{smallmatrix}\right)=\left[\begin{smallmatrix}
        \U & \V\\
        \W & \G
    \end{smallmatrix}\right] \left(\begin{smallmatrix}
        w\\
        u
    \end{smallmatrix}\right)$, which is formed by stacking the dynamics of $N$ sub-systems, where the regulated output $z=\left(\begin{smallmatrix}
        Q^{\frac{\tp}{2}} & 0
    \end{smallmatrix}\right)^\tp x + \left(\begin{smallmatrix}
        R^{\frac{\tp}{2}} & 0
    \end{smallmatrix}\right)^\tp u$ for cost matrices Q and R. Then the optimal $\K\in\Stau$ in presence of heterogeneous delays by minimizing the cost $\normm{\,\U + \V\K(I-\ \G\K)^{-1}\W\,}_2^2$ is obtained by splitting the cost into N separate convex optimization problems and handling the continuous-time delays using loop-shifting. The solution of each of the N problems is $\K_i = \Pi_{u_i}\tilde{\K}_i\left(I-\Pi_{b_i}\tilde{\K}_i\right)^{-1}$ for all $i\in\{1,\cdots,N\}$, $\Pi_{b_i}$ and $\Pi_{u_i}$ are Finite Impulse Response (FIR) blocks~{\!\!\cite[Sec.~1.5.1]{kashyap2023thesis}}, and $\tilde{\K}_i$ is the delay-free optimal LQG controller~\cite{zdg} for a rational transformation of $\Pp$.
\end{lem}
	The description of the infinite-dimensional FIR blocks is beyond the scope of this paper. See {\!\!\cite[Sec.~1.5.1]{kashyap2023thesis}} and {\!\!\cite[App.~A]{10373880}} for details. We present a brief outline for deriving the optimal decentralized wavelength controller.
\subsection{Solution outline}
We begin with the setup in~\cref{opt}. Define the control cost matrix $R\defeq \left(\begin{smallmatrix}
        \rho_P & 0\\
        0 & \rho_S
    \end{smallmatrix}\right)$ and performing Cholesky decomposition on $Q$ and $R$, we form the regulated output $z$ for wavelength control. We define the global plant 
 \begin{align*}\label{eq:ss_problem} 
 \Pp^{\lambda} &\defeq \left[\begin{smallmatrix}
 \U^{\lambda} & \V^{\lambda}\\
		\W^{\lambda} & \G^{\lambda}
 \end{smallmatrix}\right]\\
  &\defeq
	\left[\begin{array}{c|cc}
		\left(\begin{smallmatrix}
        A_{P} & 0\\
        0 & A_{S}
    \end{smallmatrix}\right) & \left(\begin{smallmatrix}
        B_{1_{P}} & 0\\
        0 & B_{1_{S}}
    \end{smallmatrix}\right) & \left(\begin{smallmatrix}
        B_{2_{P}} & 0\\
        0 & B_{2_{S}}
    \end{smallmatrix}\right) \\[2pt] \hlinet
		\left(\begin{smallmatrix}
        Q^{\frac{\tp}{2}} & 0
    \end{smallmatrix}\right)^\tp & 0 & \left(\begin{smallmatrix}
        R^{\frac{\tp}{2}} & 0
    \end{smallmatrix}\right)^\tp \\
		\left(\begin{smallmatrix}
        C_{2_{P}} & 0\\
        0 & C_{2_{S}}
    \end{smallmatrix}\right) & \left(\begin{smallmatrix}
        D_{21_{P}} & 0\\
        0 & D_{21_{S}}
    \end{smallmatrix}\right) & 0\end{array}\right]. 
\end{align*} Now re-defining the cost \cref{eq:cost_problem} into the operator theoretic framework we obtain:
\begin{equation}\label{opt_new}
\begin{aligned}
\underset{\K}{\min}
\qquad & \normm{\,\U^{\lambda} + \V^{\lambda}\K(I-\ \G^{\lambda}\K)^{-1}\W^{\lambda}\,}_{\mathcal{H}_2}^2 \\
\subject \qquad & \K \in \Stau
\text{ and $\K$ stabilizes $\Pp^{\lambda}$.}
\end{aligned}
\end{equation}

\cref{opt_new} is exactly the problem setup for \cref{lem:het_decen}. Using the block-diagonal structure of $\G^{\lambda}$, and $\forall \;\K \in \Stau$, we have $\K\G^{\lambda}\K \in \Stau$, which is the defining property of quadratic invariance~\cite{rotkowitz2005characterization}. The rest of the solution process follows the same steps as in {\!\!\cite[Thm.~43]{kashyap2023thesis}}. Briefly, we leverage the block-diagonal structure of $\W^{\lambda}$ to divide \cref{opt_new} into two smaller sub-problems. Using the loop-shifting technique~\cite{mirkin2003extraction,mirkin2011dead} iteratively, we compensate for $\tau_1$, followed by $\tau_{2}-\tau_{1}$ in both the sub-problems. This results into two non-delayed LQG synthesis problems that are solved by standard optimal control techniques~\cite{zdg}.

\subsection{Continuous-time sub-system level controller}\label{subsec:cts-sub}
\begin{figure}[ht]
	\centering
 	\includegraphics[width=\linewidth]{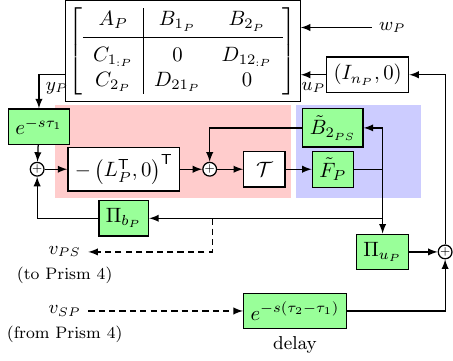}
	  \caption{The continuous-time controller for the PZT $+$ Prism $3$ is a combination of a Kalman filter (red box with gain $L_P$), a regulator (blue box with gain $\tilde{F}_P$) and two FIR blocks: $\Pi_{u_P}$ and $\Pi_{b_P}$. A correction control signal $v_{PS}$ is transmitted to the controller for Stepper $+$ Prism $4$, while a similar delayed information is received from the Stepper combination. Here $\mathcal{T}\defeq \left(sI-\left(\begin{smallmatrix}
        A_{P} & 0\\
        0 & A_{S}
    \end{smallmatrix}\right)-\left(\begin{smallmatrix}
        L_{P}C_{2_P} & 0\\
        0 & 0
    \end{smallmatrix}\right)\right)^{-1}$.}
	\label{fig:PZT_controller}
\end{figure}

The solution to \cref{opt} generates an optimal controller, which can be implemented at the sub-system level. Leveraging the existing \textit{agent-level controller} results for decentralized LQG problems {\!\!\cite[Chap.~3]{kashyap2023thesis}}, we implement a continuous-time version of our controller in \cref{fig:PZT_controller}. We introduce some notation used specifically for this implementation. $0$ is a matrix of zeros with dimensions based on context of use. $I_{n_P}$ is an identity matrix of size $n_P \times n_P$. $C_{1_{:P}}$ corresponds to the first $n_P$ columns of $\left(\begin{smallmatrix}
        Q^{\frac{\tp}{2}} & 0
    \end{smallmatrix}\right)^\tp$. Similarly $D_{{12}_{:P}}$ is the first $m_P$ columns of $\left(\begin{smallmatrix}
        R^{\frac{\tp}{2}} & 0
    \end{smallmatrix}\right)^\tp$.

\cref{fig:PZT_controller} represents a continuous-time implementation of the optimal controller for the PZT $+$ Prism $3$ sub-system. While the calculation of the Kalman gain remains the same as for a non-delayed, non-decentralized LQG problem, the delays alter LQR gain computations. Given a combination $\textup{ARE}\left(A,B,C,D\right)$ and if the Riccati assumptions hold~\cite{zdg}, there is a unique stabilizing solution $X \succ 0$ for
\begin{multline*}
    A^\tp X+XA+C^{\tp}C\\-\left(XB+C^{\tp}D\right)\left(D^{\tp}D\right)^{-1}\left(B^{\tp}X+D^{\tp}X\right)=0, 
\end{multline*} where gain $F\defeq -\left(D^{\tp}D\right)^{-1}\left(B^{\tp}X+D^{\tp}X\right)$, and $A+BF$ is Hurwitz.
So we evaluate the Kalman and LQR gains as $L_P^{\tp}=\textup{ARE}\left(A_P^{\tp},B_{1_P}^{\tp},C_{2_P}^{\tp},D_{{21}_P}^{\tp}\right)$ and $\tilde{F}_P=\textup{ARE}\left(A_P,\tilde{B}_{2_{PS}},\tilde{C}_{1_{:P}},D_{{12}_{:P}}\right)$. $\tilde{B}_{2_{PS}}$ and $\tilde{C}_{1_{:P}}$ are modifications of ${B}_{2_{P}}$, ${C}_{1_{:P}}$ and are delay-dependent~\cite{kashyap2023thesis,10373880}. The top block is a state-space matrix representation of the sub-system, which transmits wavelength measurements via the LAM. These are multiplied by the inverse of the optics gain $K_{OP}$ (not shown in \cref{fig:PZT_controller}) to generate $y_P$. However, we have a measurement delay incurred during this processing along-with any LAM delay, which are represented by the lumped delay term $\tau_1$.  Note that the number of states in the sub-controller is equal to the total number states of the overall system, i.e., $n_P + n_S$. The Kalman filter (red) predicts the states of both the sub-systems based on available information. For the PZT $+$ Prism $3$, the available information are $\{e^{-s\tau_1}x_P,\; e^{-s\tau_2}x_S\}$. The signal $v_{SP}$ already contains delayed (by $\tau_1$ s) information regarding the states of the Stepper $+$ Prism $4$ sub-system and is further delayed by a $\left(\tau_2-\tau_1\right)$ s of transmission time. The FIR blocks $\Pi_{u_P}$ (feed-forward) and $\Pi_{b_P}$ (feedback) together compensate for the continuous time-delays, without any Pad\'e approximations.

\section{Discussion}\label{sec:discussion}
In this section, we consider a discrete-level implementation of the controller presented in \cref{subsec:cts-sub} because of the pulse-by-pulse nature of the light source. A straightforward approach is to discretize the continuous-time controller, based on the sampling rate. The analog controller in \cref{fig:PZT_controller} can be implemented as a discrete controller for a given sampling period $h$ (that satisfies the Nyquist criterion) using standard digital control emulation techniques, including bilinear transformation, matched pole-zero transform, impulse variance, ZOH, First Order Hold~\cite{franklin1998digital,isermann2013digital,chen2012optimal}.

\subsection{Finite horizon cost}
\label{subsec:finite}
While the controller works for a infinite horizon problem, the light sources seldom produce bursts over an infinite horizon. A finite horizon cost defined as 
\begin{align*}\label{eq:cost_problem_finite}
J_{\textup{finite}}\defeq\E_{w} \sum_{0}^{T} \Bigl(Q\normm{x}^{2}_{2} + \sum_{i\in \{P,S\}} \rho_i \normm{u_i}^2_2\Bigr),
\end{align*} makes sense in such a case. The results from the infinite-horizon case are transferable. However the gains are no longer time-invariant and we can use dynamic programming approaches to solve the Hamiltonian-Jacobi-Bellman (HJB) equation associated with the differential Riccati equations to obtain $L_{P}(t)$ and $\tilde{F}_{P}(t)$. 

\subsection{Discrete-time decentralized control}
An alternate approach to the above wavelength control approach is to begin with a discretized version of the plant itself. For  sampling period $h$ satisfying the Nyquist criterion, a straightforward transformation of the plant dynamics occurs: $A_i\mapsto e^{A_i h}$, $B_{2_{ii}}\mapsto A_i^{-1}(e^{A_i h}-I)B_{2_{ii}}$ as $A_i$ is non-singular, $B_{1_{ii}}B_{1_{ii}}^{\tp}\mapsto \int_{\tau=0}^{h}e^{A\tau}B_{1_{ii}}B_{1_{ii}}^{\tp}e^{A^{\tp}\tau} \mathrm{d}\tau$, $D_{{21}_{ii}}D_{{21}_{ii}}^{\tp}\mapsto \frac{D_{{21}_{ii}}D_{{21}_{ii}}^{\tp}}{h}$ for $i\in\{P,S\}$, with rest of the terms remaining the same. The corresponding Kalman and LQR gains are obtained by solving a discrete ARE for the infinite horizon cost case. Akin to \cref{subsec:finite}, we can solve for time-varying gains using the Bellman equation for the finite-horizon cost for discrete decentralized LQG problem~\cite{nayyar}. This approach loses the elegance and exactness of handling continuous time-delays due to the underlying approximation involved. However the structure of the discrete implementation remains similar to \cref{fig:PZT_controller} with a discrete Kalman filter, discrete LQR, finite-dimensional FIR blocks, and time delays. 
\begin{figure}[ht]
	\centering
 	\includegraphics[width=\linewidth]{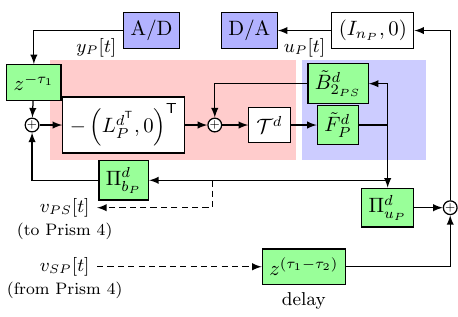}
	  \caption{The discrete-time controller for the PZT $+$ Prism $3$ is a combination of a discrete Kalman filter (red box with gain $L_P^d$), a regulator (blue box with gain $\tilde{F}_P^d$) and two discrete FIR blocks: $\Pi_{u_P}^d$ and $\Pi_{b_P}^d$. Here $\mathcal{T}^d\defeq \left(zI-\left(\begin{smallmatrix}
        e^{A_P h} & 0\\
        0 & e^{A_S h}
    \end{smallmatrix}\right)-\left(\begin{smallmatrix}
        L_{P}^d C_{2_P} & 0\\
        0 & 0
    \end{smallmatrix}\right)\right)^{-1}.$ The superscript $d$ refers to discrete-time versions of corresponding matrices.}
	\label{fig:PZT_controller_discrete}
\end{figure}

\subsection{Optimality of performance}
Here we show that the controller architectures provided in \cref{fig:PZT_controller,fig:PZT_controller_discrete} incur the lowest cost in comparison to existing approaches for wavelength control. Current control techniques utilize a discontinuous approach, where the coarse actuator serves to desaturate the finer actuator~\cite{kevin_tutorial}. Further the amount of wavelength target change determines whether a single actuator or a multi-actuator combination is used for control~\cite{kevin_tutorial}. The decentralized control method allows for synchronization of the control action of both the sub-systems, ensuring a more continuous level of control. We define the average cost incurred by any sub-optimal strategies as $J_{\textup{dec,del}}+\Delta$, where $J_{\textup{dec,del}}$ is the average optimal cost for a global controller $\K_{\textup{dec,del}}\defeq\left(\begin{smallmatrix}
        e^{-s\tau_1}\K_{P} & 0\\
        0 & e^{-s\tau_1}\K_{S}
    \end{smallmatrix}\right)$ given the cost function defined in  \cref{eq:cost_problem}, and $\Delta$ is the additional cost due to sub-optimal implementations: for instance, open-loop control of the stepper. Note that the measurement delay $\tau_1$ is already considered in this framework. Defining the cost $J_{\textup{cen,del}}$ for $\left(\begin{smallmatrix}
        e^{-s\tau_1}\K_{P} & e^{-s\tau_1}\K_{SP}\\
        e^{-s\tau_1}\K_{PS} & e^{-s\tau_1}\K_{S}
    \end{smallmatrix}\right)$, we have  $J_{\textup{cen,del}}<J_{\textup{dec,del}}$ from results in~{\!\!\cite[Thm.~17]{10373880}} and~{\!\!\cite[Lem.~9]{mirkin2003every}}. The final step is to establish that the cost $J_{\textup{cen},\tau_2}$ defined for $\left(\begin{smallmatrix}
        e^{-s\tau_1}\K_{P} & e^{-s\tau_2}\K_{SP}\\
        e^{-s\tau_2}\K_{PS} & e^{-s\tau_1}\K_{S}
    \end{smallmatrix}\right)$ satisfies $J_{\textup{cen,del}}< J_{\textup{cen},\tau_2} < J_{\textup{dec,del}}$. While $J_{\textup{cen,del}}< J_{\textup{cen},\tau_2}$ is implicit from~{\!\!\cite[Thm.~17]{10373880}}. Using a simple limit argument and $J_{\textup{cen},\tau_2}$ being a monotonic non-decreasing function with increasing delays~{\!\!\cite[Prop.~6]{MIRKIN20121722}}, we establish $J_{\textup{cen},\tau_2} < J_{\textup{dec,del}}$. Indeed for $\lim_{\left(\tau_2-\tau_1\right) \to \infty}$, $\K \to \K_{\textup{dec,del}},$ and $J_{\textup{cen},\tau_2} \to J_{\textup{dec,del}}$. Thus $J_{\textup{cen},\tau_2}<J_{\textup{dec,del}}+\Delta$ establishes the `optimal' performance of the decentralized controllers in comparison to existing approaches.

    \subsection{Aliased disturbances}
    The light source open loop frequency data has several narrow-band, periodic components at known frequencies, some of which are aliased~\cite{riggs2012rejection}. One of the primary sources of such wavelength disturbances are blowers in the gas chamber that cause acoustic waves inside the chamber, which couple into the optics~\cite{kevin_tutorial}. For a given sampling rate, periodic disturbances can be modeled as sinusoidal signals, and this disturbance model can be integrated with the state-space formulation in \cref{eq:state_space_2,eq:state_space_plant} as part of the PZT $+$ Prism $3$ sub-system. However, this could induce the loss of unobservability for the estimation and uncontrollability for the regulator problems in the augmented state-space system. This is not a hard constraint for solving AREs in the finite-horizon case; while workarounds exist for the infinite-horizon case. We can add a small, non-zero value to disturbance matrices for $B_2$ and $C_2$ to satisfy observability and controllability. With the prior changes, we can solve the optimal decentralized control problem for the augmented system, while simultaneously compensating for these periodic disturbances. Alternately, there exists a rich literature on implementable disturbance cancellation techniques, which are beyond the scope of this paper.

The discrete-time formulations of \cref{fig:PZT_controller,fig:PZT_controller_discrete} consider a constant sampling rate. However practical considerations could require a variable sampling rate, due to laser feature specifications or the sampler characteristics. Further hardware or other restrictions could require a sampling rate different from the control rate~\cite{riggs2012rejection}. However the elegant observer-regulator separation structure for the decentralized control problem (red-blue box separation) allows for implementation of modified Kalman filters and modified LQRs to handle these variations. It is worth noting that the LQR gain is solely delay-dependent, with only certain elements of this matrix being functions of $\tau_1$ and $\tau_2-\tau_1$. This allows for faster tuning in case of time-varying delays.
\section{Conclusions}
\label{sec:conclufuture}
We studied a practical example of a structured optimal control problem where multiple optics along-with their actuators,
communicate over a delayed network for wavelength control of light in pulsed light sources. Specifically, we characterized the efficient implementation of optimal controllers at the individual optics level. We showed that these controllers incur the minimum cost compared to existing control approaches used for wavelength control in lithography. We also discussed several approaches to handle practical constraints including disturbances, intermittent sampling rates, and finite-horizon cost specifications for these controllers.

While this paper only considered the case of two prisms with their corresponding actuators, our approach is generalizable to multiple optics and actuator combinations to control wavelength in light sources used for photolithography.
\if\MODE3
\bibliographystyle{IEEEtran}
\bibliography{rdlqr}
\else
\bibliographystyle{abbrv}
{\small \bibliography{rdlqr}}
\fi

\end{document}